# Voltage-Controlled Phosphate Precipitation Gating in Solid-State Nanopore Memristors

Weihong Wang[a], German Lanzavecchia[b,c], Ali Douaki[b,c], Shukun Weng[b], Yanqiu Zou[d], Huaizhou Jin[e], Alberto Giacomello[f], Lorenzo Iannetti[f], Roman Krahne[b], Shangzhong Jin[a*], Makusu Tsutsui[*g], and Denis Garoli[a,b,c*]

a. *College of Optical and Electronic Technology, China Jiliang University, Hangzhou 310018, China*

b. *Istituto Italiano di Tecnologia, Via Morego 30, 16136 Genova, Italy.*

c. *Dip. di Scienze e Metodi dell'Ingegneria, Università di Modena e Reggio Emilia, via Amendola 2, 42122 Reggio Emilia, Italy.*

d. *State Key Laboratory of Modern Optical Instrumentation, College of Optical Science and Engineering, Zhejiang University, Hangzhou 310027, China.*

e. *Key Laboratory of Quantum Precision Measurement, College of Physics, Zhejiang University of Technology, Hangzhou, China.*

f. *Department of Mechanical and Aerospace Engineering, Sapienza University of Rome, Via Eudossiana 18, 00184 Rome, Italy.*

g. *SANKEN, The University of Osaka, Osaka, Ibaraki 567-0047, Japan*

Corresponding authors: Prof. Denis Garoli – denis.garoli@unimore.it; Prof. Makusu Tsutsui - makusu32@sanken.osaka-u.ac.jp; Prof. Shangzhong Jin - jinsz@cjlu.edu.cn

# Abstract

Nanofluidic memristors preserve a record of electrical activity via ion migration and alterations in conductance that depend on the history of the device's state. These characteristics make them suitable for aqueous, energy-efficient, and biologically compatible neuromorphic systems. To establish the viability of fluidic memristors for mimicking the brain’s dynamic behavior, a more thorough understanding of the memristive materials and the underlying switching processes is required. In this study, we systematically examined a recently introduced memristive device based on in-pore chemical reactions, where the combined influence of electrolyte composition and pore architecture on precipitation-gated memory remains poorly understood. To address this, we constructed an asymmetric electrochemical system using $CaCl_2$ and phosphate solutions separated by $SiN_x$ solid-state nanopores. We explored how variations in pH, phosphate concentration, pore geometry, and voltage pulsing regimens affect the electrical characteristics and memristive performance. Comparison of the single pore and the 3 × 3 array showed that parallel pores produced smoother pH- and concentration-dependent hysteresis and pulse responses, whereas the single pore retained larger, nonmonotonic changes.

## Introduction

Nanofluidic memristors record past inputs through ion transport rather than electron transfer [1-6]. Their state variables include ion concentration, interfacial charge, confined reactions, and effective pore size [7-9]. These variables support aqueous, low-power devices for ionic neuromorphic computing, chemical sensing, and information processing. Reported mechanisms include ion concentration polarization[10, 11], mechanical deformation[12], and nanoscale electrochemical reactions [12, 13]. Despite this diversity, device fabrication and control of memristive dynamics remain challenging [14, 15].

In particular, one central question in nanofluidic memristors is how tunable physical and chemical variables control I-V hysteresis and pulse responses. Several ion-transport processes can trigger nanofluidic memristive behavior[16-18]. Voltage-driven ion enrichment and depletion generate frequency-dependent pinched hysteresis in conical nanopores[19], charged nanochannels, and two-dimensional confined channels. These systems show history-dependent conductance. Porous membranes and nanopore arrays place many pores in parallel and convert single-pore responses into larger array currents[20-23]. Voltage-driven concentration polarization can also push the local ion product above the precipitation threshold of a sparingly soluble salt. This process can cause nanoscale precipitation, current oscillations, negative differential resistance, and reversible current blockage[24-30]. Thus, a nanopore can act as both an ion channel and a voltage-controlled reaction space. Precipitation and dissolution inside a pore offer a direct route to chemically tunable nanofluidic memristors[29]. We have recently reported studies on reversible precipitation of sparingly soluble salts such as $CaHPO_4$ inside nanopores. The in-pore reaction mechanism allows a memory effect on the ionic conductance[31]. Salt concentration, scan rate, and voltage amplitude altered the hysteresis area, threshold voltage, and pulse response. Voltage-driven deposition and redissolution of metal phosphates also produced strong rectification and memristive responses in $SiN_x$ nanopores[31-33]. These studies, however, focused on demonstrating precipitation gating. Systematic comparisons of reaction conditions and pore geometry were not reported. The separate roles of pH, phosphate supply, parallel pores, and pore area therefore remain unclear[34].

These factors may control gating strength, reversibility, and temporal accumulation. Here, we built an asymmetric $CaCl_2$/phosphate electrolyte system across $SiN_x$ solid-state nanopores. We compared current responses across pH, phosphate concentration, pore geometry, and pulse excitation protocols. We compared two main devices: (i) a single pore with a nominal pore size of $\phi$ =250 nm and (ii) a 3 × 3 array of nine pores, each approximately 250 nm in diameter. I-V hysteresis quantified history dependence during voltage sweeps. Pulse sampling tested whether a conductance state remained readable at a fixed voltage[22, 23, 35, 36]. We included an $\phi$ = 150 nm single pore only as an auxiliary geometry control; the Supporting Information reports its response data. These comparisons clarify how pH and phosphate supply regulate precipitation-gated memory. They also guide pore design and experimental settings for chemically tunable solid-state nanopore memristors.

## Results and Discussion

Figure 1 shows the primary single-pore and 3 × 3 nanopores array devices. We tested both in the same two-reservoir configuration. Supplementary Figs. S1 and S2 report additional data for the auxiliary $\phi$ = 150 nm single pore.

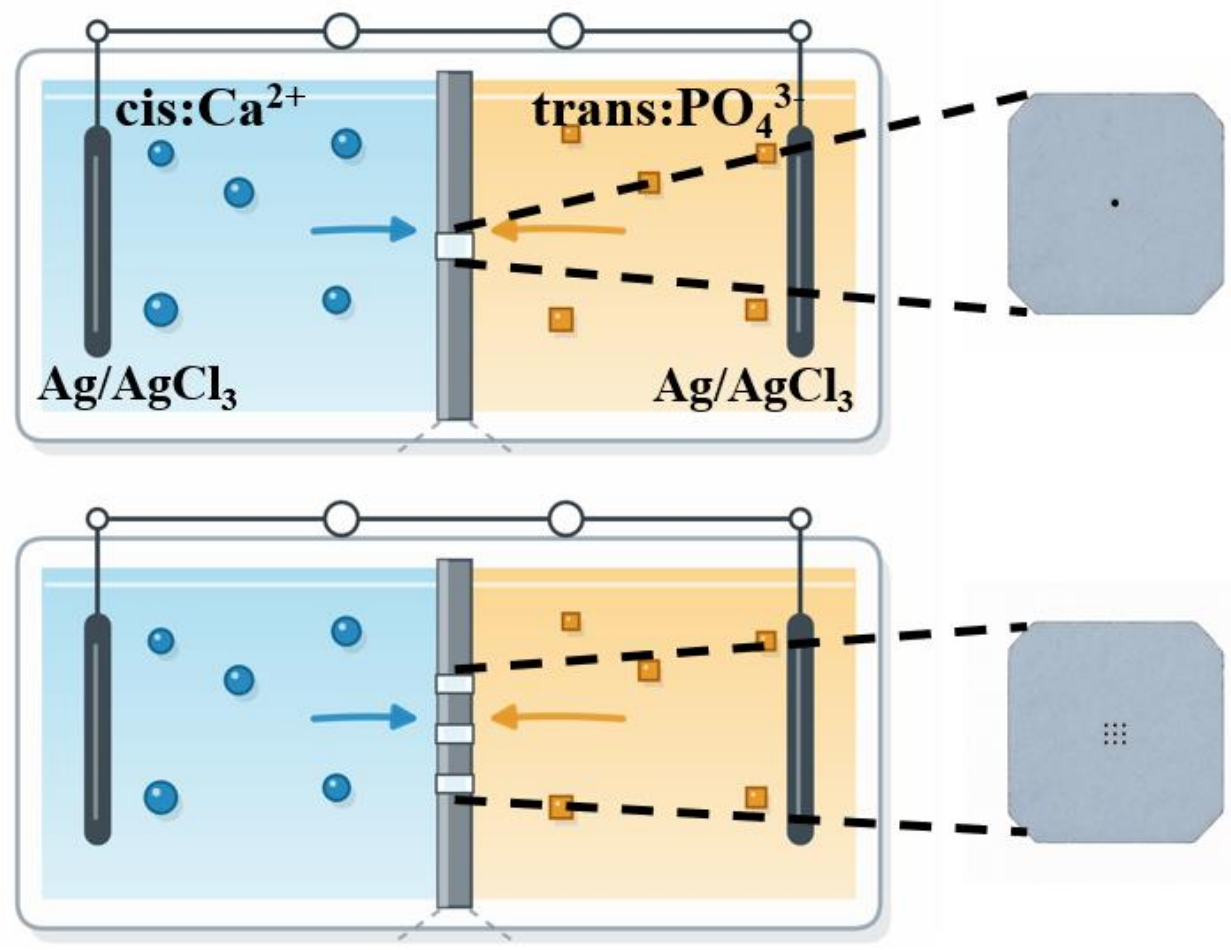

**Fig. 1**. Nanopore devices and measurement geometry. (a) Single pore. (b) 3 × 3 nanopores array of pores with 200 nm center-to-center spacing. The cis electrode was held at 0 V, and voltage is reported as $V = V_{trans} - V_{cis}$. Negative and positive bias denote the open and closed states, respectively.

pH changes phosphate speciation and the supersaturation available for calcium-phosphate formation. Increasing pH shifts the distribution from $H_2PO_4^-$ toward $HPO_4^{2-}$ and $PO_4^{3-}$. Several calcium-phosphate phases are possible, but the electrical measurements do not identify a specific product. Equations 1–3 therefore describe phosphate acid–base equilibria, not a phase assignment. Figure 2a summarizes the proposed precipitation–dissolution cycle, including pore sealing, precipitate dissolution, pore reopening, and renewed precipitation.

$$H_3PO_4 \rightleftharpoons H^+ + H_2PO_4^- \quad (1)$$

$$H_2PO_4^- \rightleftharpoons H^+ + HPO_4^{2-} \quad (2)$$

$$HPO_4^{2-} \rightleftharpoons H^+ + PO_4^{3-} \quad (3)$$

The nanopores array showed a comparatively smooth pH-dependent response (Fig. 2b,d). We compared hysteresis using the dimensionless normalized loop area, $A_{norm}$, together with the absolute loop area, $A_{hys}$ (see methods for details).

$A_{norm}$ was 0.0986 at pH 5.5 and 0.0992 at pH 6.5, and decreased to 0.0385 at pH 9.5. Here, $G_{d,max}^{\pm}$ denotes the maximum local differential conductance, *dI/dV*, within the corresponding bias regime. $G_{fit}^{\pm}$ denotes the slope obtained by linearly fitting the high-bias segment of the branch-averaged I–V curve. Bias-resolved analysis showed that the maximum negative-bias differential conductance, $G_{d,max}^{-}$, remained between 207.5 and 231.3 nS from pH 5.5 to 8.5 and increased to 337.9 nS at pH 9.5 (Supplementary Fig. S7a). The high-bias fitted conductance, $G_{fit}^{-}$, likewise increased overall from 122.0 nS at pH 5.5 to 262.7 nS at pH 9.5. Thus, the negative-bias open-state conductance did not increase as pH decreased. The observed trend may reflect changes in phosphate speciation, ionic strength, and precipitation or dissolution; the electrical data alone do not separate these contributions. In the positive-bias closed state, $G_{d,max}^{+}$ varied from 51.1 to 82.4 nS and $G_{fit}^{+}$ from 7.6 to 17.7 nS, remaining well below the negative-bias conductance (Supplementary Fig. S7c). This supports persistence of a low-conductance state during the sweep, but does not establish long-term precipitate stability[28-30, 33].

The single pore was more sensitive and nonmonotonic (Fig. 2c,d). $A_{norm}$ reached 0.2937 at pH 8.5 but was 0.1451 at pH 7.5. Its $G_{d,max}^{-}$ values were 28.7, 62.4, 53.9, 25.8, and 31.2 nS from pH 5.5 to 9.5, respectively, whereas $G_{d,max}^{+}$ remained between 8.3 and 14.5 nS (Supplementary Fig. S7). The nonmonotonic negative-bias response shows that a local pore does not follow the more regular trajectory observed in the nanopores array. Scan-rate measurements at 0.04, 0.06, and 0.12 V $s^{-1}$ further separated the two devices (Fig. 2e–g). For the nanopores array, $A_{norm}$ increased from 0.0026 at 0.04 V $s^{-1}$ to 0.0857 at 0.12 V $s^{-1}$, whereas the single pore reached its maximum at 0.06 V $s^{-1}$ and decreased at 0.12 V $s^{-1}$. These data demonstrate different scan-rate dependences over the tested range, rather than independently measured characteristic response times[6, 7, 18, 23, 37]. For clarity of presentation, the single-pore pH loops in Fig. 2c were 0V capacitance-corrected; the uncorrected curves are provided in Supplementary Fig. S5a. hysteresis areas versus scan rate.

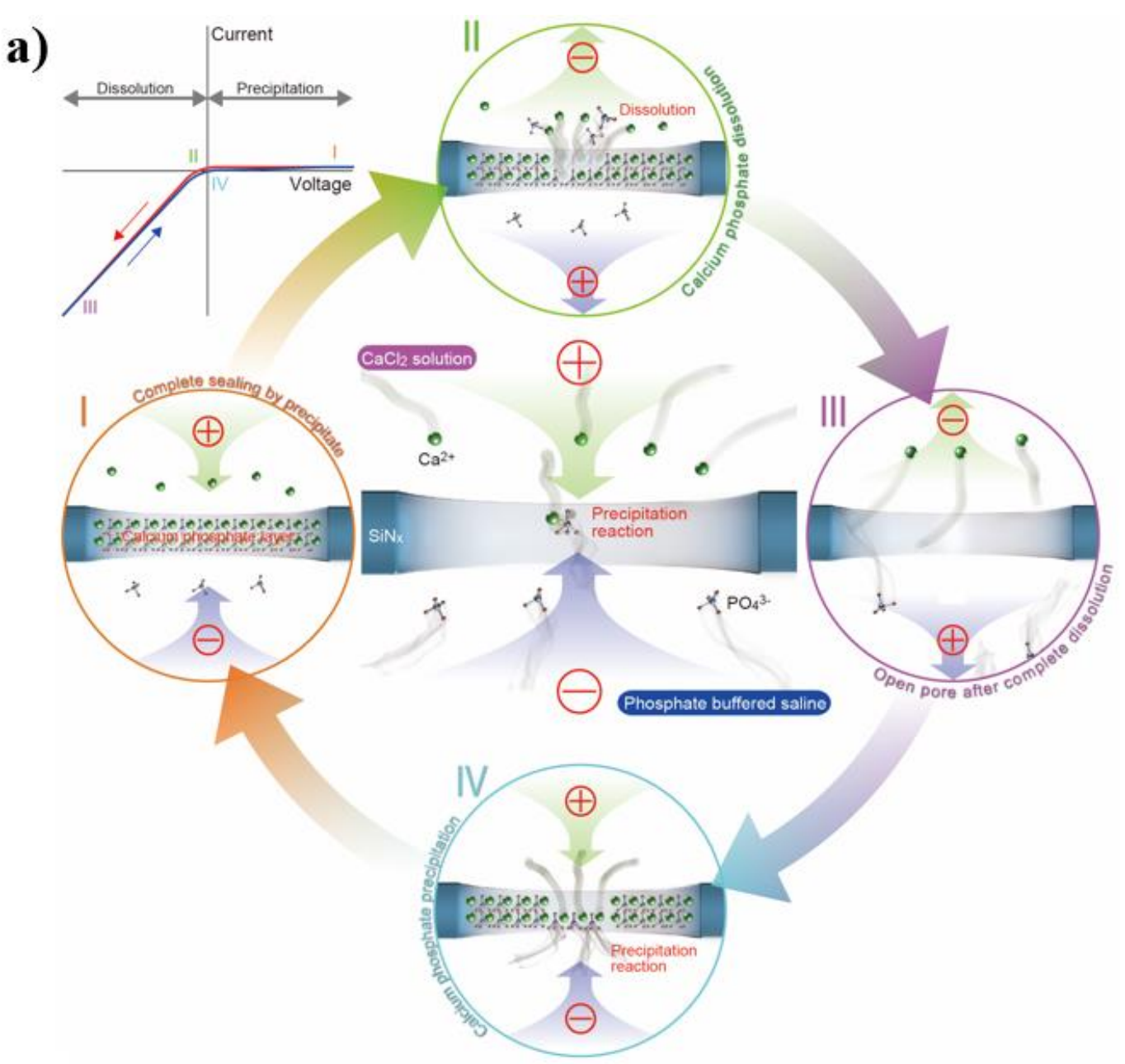

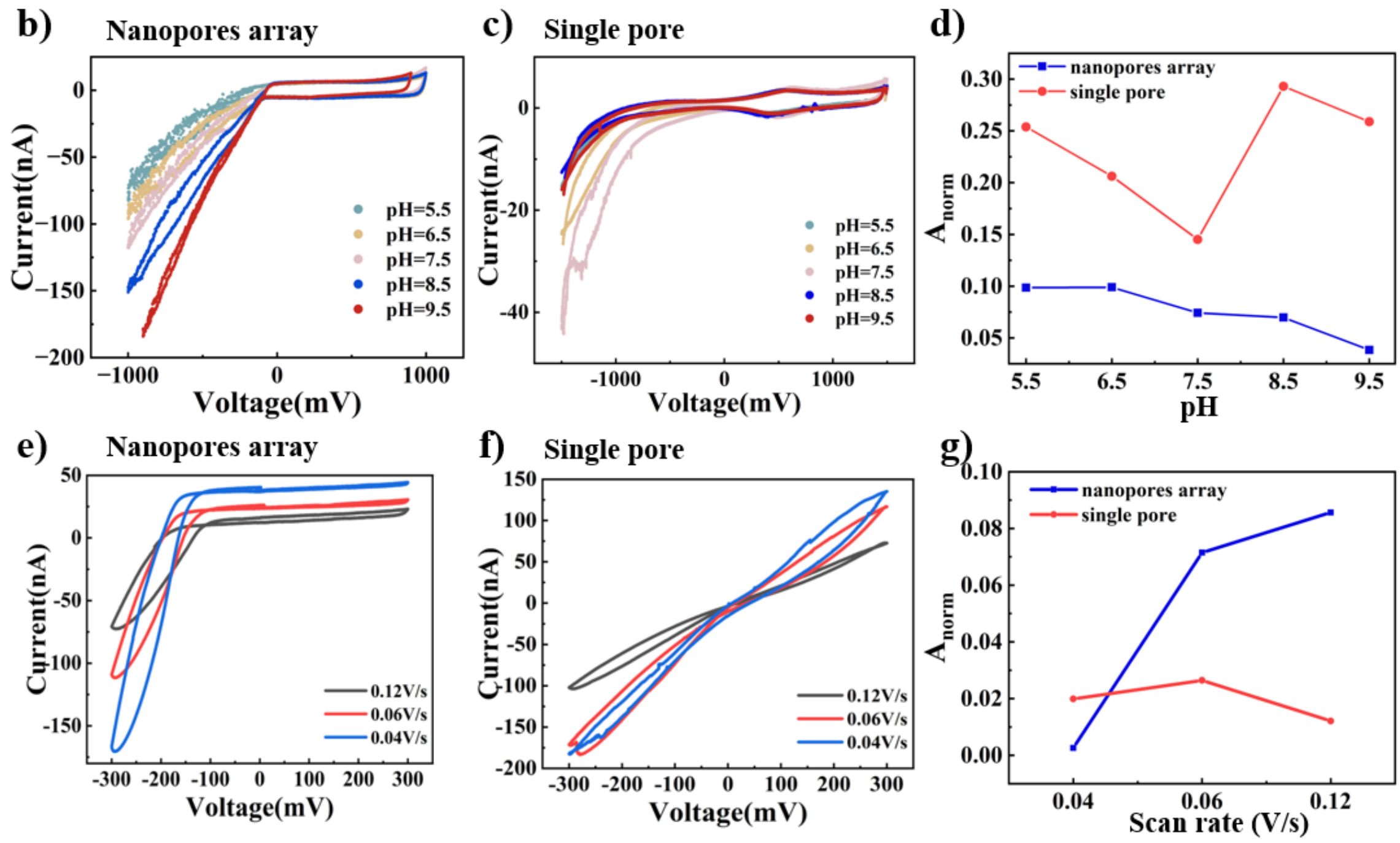

**Fig. 2.** Proposed precipitation–dissolution mechanism and pH-dependent I–V hysteresis. (a) Proposed four-stage cycle comprising calcium-phosphate precipitation and pore sealing (I), precipitate dissolution (II), pore reopening after dissolution (III), and renewed precipitation (IV). (b) I–V loops from pH 5.5 to 9.5 for the 3 × 3 nanopore array. (c) Baseline-corrected I–V loops for the r = 250 nm single pore; the uncorrected curves are shown in Supplementary Fig. S5a. (d) $A_{norm}$ versus pH. (e,f) I–V loops at 0.04, 0.06, and 0.12 V $s^{-1}$ for the array and single pore at pH 8.5. (g) $A_{norm}$ versus scan rate. Supplementary Fig. S6a shows the corresponding absolute hysteresis area, $A_{hys}$, versus scan rate. Supplementary Fig. S1a shows the corresponding $\phi$ = 150 nm single-pore control, and Supplementary Figs. S7 and S9 provide the bias-resolved conductance and apparent-capacitance analyses.

Phosphate concentration sets the reactant supply to the confined reaction zone. Voltage brings $Ca^{2+}$ and phosphate species together inside the pore; when their local ion product exceeds the relevant solubility limit, a calcium-phosphate solid can reduce the conducting cross-section[24-26]. Equations 4–6 list candidate equilibria rather than identified products.

$$Ca^{2+} + HPO_4^{2-} \rightleftharpoons CaHPO_4(s) \quad (4)$$

$$3Ca^{2+} + 2HPO_4^{2-} \rightleftharpoons Ca_3(PO_4)_2(s) + 2H^+ \quad (5)$$

$$5Ca^{2+} + 3HPO_4^{2-} + H_2O \rightleftharpoons Ca_5(PO_4)_3OH(s) + 4H^+ \quad (6)$$

The single pore showed a concentration-dependent but nonmonotonic increase in hysteresis (Fig. 3b,c). $A_{norm}$ was 0.0342 at 1 mM and 0.0308 at 5 mM, increased to 0.0843 at 10 mM, and reached 0.1008 at 100 mM. Bias-resolved analysis revealed a threshold-like change between 5 and 10 mM: $G_{d,max}^-$ increased from 83.9 to 222.7 nS and remained between 232.1 and 244.0 nS at 50–100 mM (Supplementary Fig. S8a). The positive-bias conductance was also nonmonotonic, with $G_{d,max}^+$ ranging from 83.1 to 236.0 nS. The change between 5 and 10 mM therefore marks a condition-dependent onset in this device, not a universal precipitation threshold. For presentation, the single-pore concentration loops in Fig. 3b were 0V capacitance-corrected (see methods section); the uncorrected curves are provided in Supplementary Fig. S5b.

The array response was smoother across concentration (Fig. 3a,c). $G_{d,max}^-$ decreased from 800.2 nS at 1 mM to 633.0 nS at 10 mM, increased locally to 705.1 nS at 50 mM, and was 653.4 nS at 100 mM (Supplementary Fig. S8a). The stronger concentration dependence occurred in the positive-bias closed state: $G_{d,max}^{+}$ decreased from 438.6 nS at 1 mM to 196.8 nS at 100 mM, a 55% change compared with an 18% endpoint change in $G_{d,max}^{-}$. Under the polarity used here, this positive-bias dependence is consistent with concentration-dependent formation or persistence of calcium-phosphate blockage and changes in residual mobile-ion concentration. It should not be assigned solely to $Ca^{2+}$ carrier concentration. Previous work established that the conducting ion and in-pore reaction depend on voltage polarity[33]; the polarity of the open and closed states in the present experiment is the reverse of that previous configuration. At 100 mM, $A_{norm}$ was 0.0681 for the array and 0.1008 for the single pore. The array therefore reduced condition-to-condition variation at the cost of some local response amplitude. Supplementary Fig. S6b reports the corresponding absolute hysteresis areas versus scan rate.

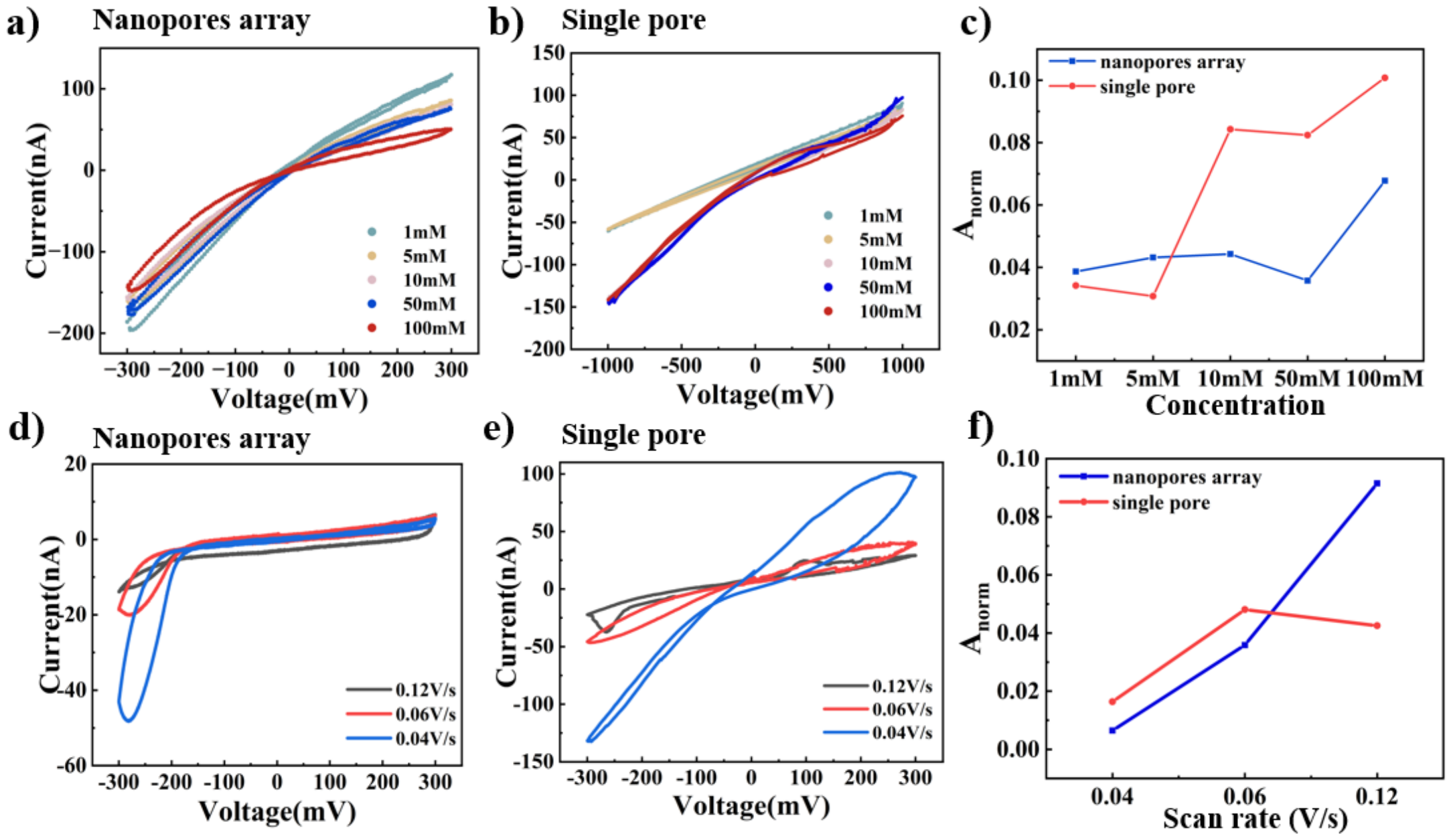

Fig. 3. $Na_2HPO_4$ concentration and scan rate define different response windows in parallel and single pores. (a) I–V loops from 1 to 100 mM for the 3 × 3 nanopore array. (b) 0V capacitance-corrected I–V loops for the single pore; the uncorrected curves are shown in Supplementary Fig. S5b. (c) $A_{norm}$ versus concentration. (d,e) I–V loops at 0.04, 0.06, and 0.12 V $s^{-1}$ for the nanopore array and single pore at 100 mM $Na_2HPO_4$. (f) $A_{norm}$ versus scan rate. Supplementary Fig. S6b shows the corresponding absolute hysteresis area, $A_{hys}$, versus scan rate. Supplementary Fig. S1b shows the corresponding $\phi$ = 150 nm single-pore control, and Supplementary Figs. S8 and S9 provide bias-resolved current, conductance, and apparent-capacitance analyses.

The concentration data argue against a single monotonic optimum. Insufficient phosphate supply limits deposition, whereas high supply may promote persistent blockage or residual nuclei that alter recovery[27-29, 33]. The single pore exposes this competition more strongly; the nanopores array averages it across parallel pathways. A favorable chemical window must therefore be defined jointly by hysteresis, pulse readability, and reversibility.

The pinched I–V loops provide the first measure of precipitation-gated memory, whereas pulse sampling tests whether a written conductance state remains readable at a fixed voltage [22, 23, 35, 36]. We therefore applied positive and negative programming pulses across the pH and $Na_2HPO_4$ series (Figs. 4 and 5). These sequences probed conductance changes associated with precipitation and recovery without assuming complete dissolution or cycle-to-cycle reproducibility. Figures 4 and 5 show the full 300-pulse records for the nanopores array and single pore across all available conditions. Supplementary Figs. S3 and S4 retain the original 150-pulse views, and Supplementary Fig. S2 reports the corresponding $\phi$ = 150 nm control trajectories.

The pH pulse series remained condition dependent (Fig. 4). We define $\Delta I = I_{end} - I_0$, where $I_0$ is the median of the first five readouts and $I_{end}$ is the median of the final 20% of readouts. Thus, positive and negative ΔI denote higher and lower endpoint currents, respectively. Under the ±60 mV programming window, array ΔI changed from −0.5005 nA at pH 5.5 to +1.1322 nA at pH 6.5 (Table 1). More acidic conditions may accelerate precipitate dissolution during negative-bias reopening, but the sign changes across pH and programming window show that this effect is not monotonic. The array produced a more regular aggregate response in selected windows, whereas the single pore retained a more complex pH dependence. Supplementary Fig. S3 shows the original 150-pulse view of this pH series.

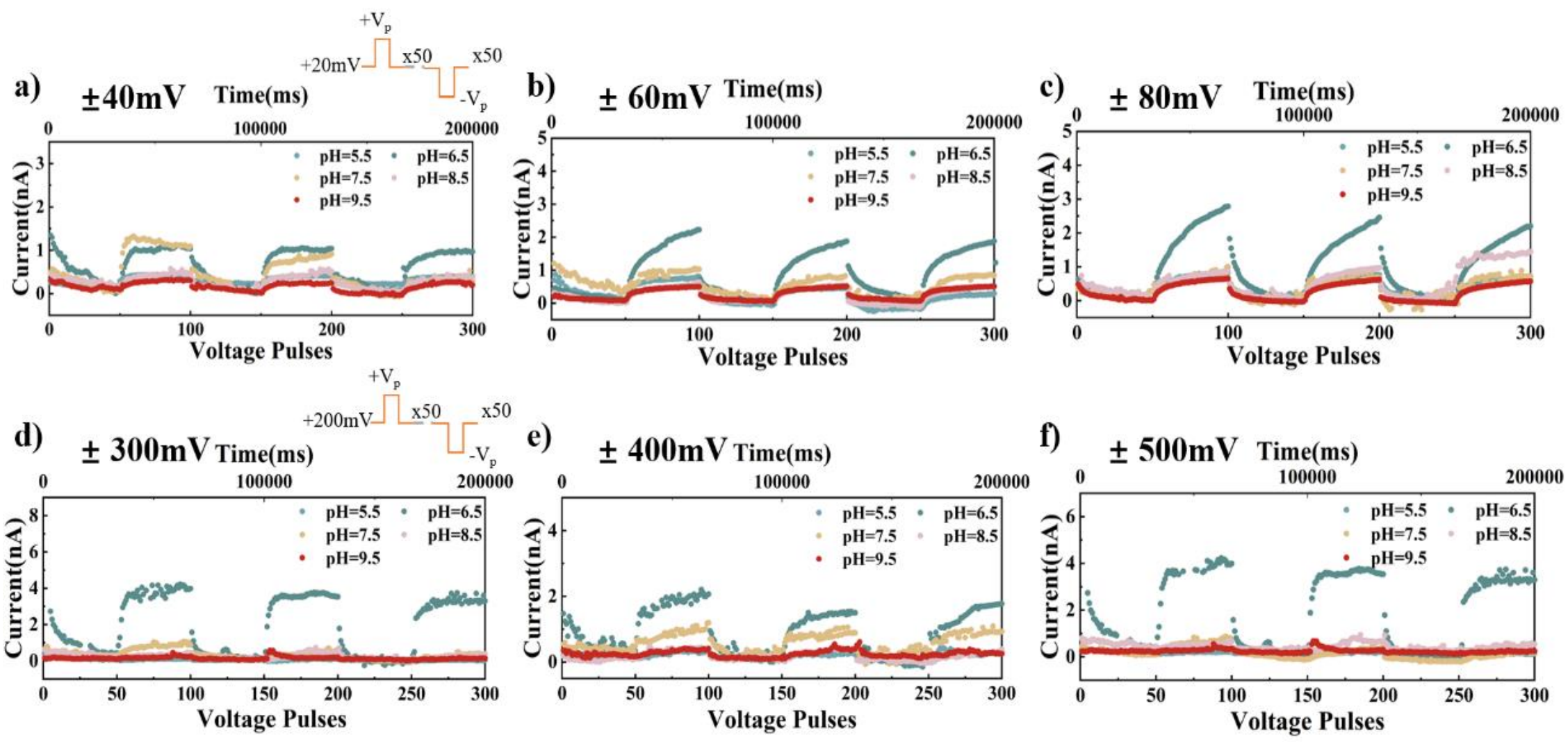

**Fig. 4.** Sampled-current trajectories across pH, pore geometry, and programming voltage window. (a–c) The 3 × 3 nanopores array at ±40, ±60, and ±80 mV. (d–f) The single pore at ±300, ±400, and ±500 mV. The 300-pulse records contain six consecutive 50-pulse blocks under alternating programming polarities. The read voltage was +20 mV for the nanopores array and +200 mV for the single pore. Supplementary Fig. S3 retains the original 150-pulse view; Supplementary Fig. S2a–c shows the corresponding $\phi$ = 150 nm single-pore control.

**Table 1.** Descriptive endpoint current change, ΔI (nA), across pH.

| voltage | **pH = 5.5** | **pH = 6.5** | **pH = 7.5** | **pH = 8.5** | **pH = 9.5** |
|---|---|---|---|---|---|
| ±40 mV (pore array) | 0.1305 | -0.2604 | -0.2411 | -0.2808 | -0.1770 |
| ±60 mV (pore array) | -0.5005 | 1.1322 | -0.3890 | 0.0991 | 0.2289 |
| ±80 mV (pore array) | 0.1923 | 0.9247 | 0.3801 | -0.0702 | 0.0763 |
| ±300 mV (single pore) | -0.1831 | -1.3169 | -0.4028 | -0.2686 | -0.0366 |
| ±400 mV (single pore) | 0.0021 | -0.0045 | 0.3801 | 0.0305 | -0.0681 |
| ±500 mV (single pore) | -0.1831 | -0.8911 | -0.4028 | -0.3357 | -0.6939 |

The single-pore measurements used larger programming windows (±300–500 mV versus ±40–80 mV for the array) and a tenfold higher read voltage to obtain a measurable signal from one conduction pathway. The larger single-pore current excursions therefore reflect both stronger electrical drive and local opening–closing events. They should not be interpreted as an intrinsically larger geometry-normalized response. Because ΔI is an endpoint metric from a single trajectory, it does not quantify device lifetime or replicate-to-replicate variability.

The concentration pulse series is consistent with a smoother array-level response, although the different programming and read voltages prevent attributing this contrast to pore number alone (Fig. 5). At 100 mM $Na_2HPO_4$, nanopores array ΔI remained positive and tightly grouped at 2.600, 2.664, and 2.858 nA across the ±40, ±60, and ±80 mV programming windows (Table 2). Sufficient phosphate activity may help replenish reactive ions locally depleted by precipitation during the pulse train, thereby sustaining the endpoint current. At 200 mM, all three array values also remained positive (1.286, 1.064, and 1.534 nA), although they were lower than at 100 mM. Thus, excess phosphate did not abolish the memristive endpoint response, but neither did it monotonically enhance its magnitude. The single pore gave 2.625, −0.268, and 1.123 nA at 100 mM, including a sign reversal, consistent with a less regular local response. Supplementary Fig. S4 shows the original 150-pulse view of this concentration series.

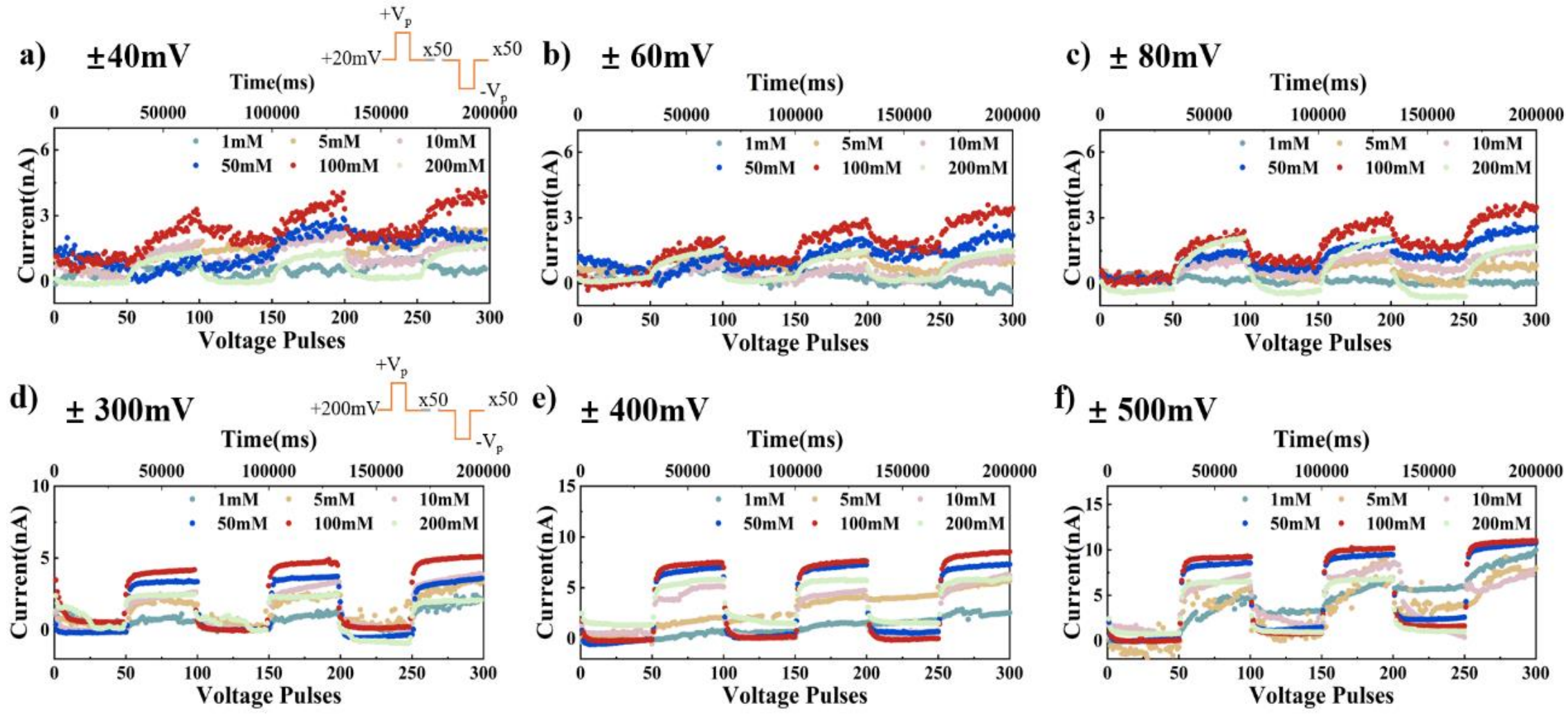

**Fig. 5**. Sampled-current trajectories across $Na_2HPO_4$ concentration, pore geometry, and programming voltage window. (a–c) The 3 × 3 nanopores array at ±40, ±60, and ±80 mV. (d–f) The single pore at ±300, ±400, and ±500 mV. Supplementary Fig. S4 retains the original 150-pulse view; Supplementary Fig. S2d–f shows the corresponding ϕ = 150 nm single-pore control.

Table 2. Descriptive endpoint current change, ΔI (nA), across $Na_2HPO_4$ concentration.

| voltage | **1 mM** | **5 mM** | **10 mM** | **50 mM** | **100 mM** | **200 mM** |
|---|---|---|---|---|---|---|
| ±40 mV (pore array) | 0.570 | 1.298 | 0.764 | 0.740 | 2.600 | 1.286 |
| ±60 mV (pore array) | -0.488 | 0.316 | 0.663 | 0.595 | 2.664 | 1.064 |
| ±80 mV (pore array) | -0.207 | 0.363 | 1.053 | 1.882 | 2.858 | 1.534 |
| ±300 mV (single pore) | 0.262 | 1.960 | 2.939 | 3.328 | 2.625 | 0.333 |
| ±400 mV (single pore) | 2.110 | 5.155 | 0.068 | 1.057 | -0.268 | -0.171 |
| ±500 mV (single pore) | 5.695 | 4.254 | 0.245 | 2.130 | 1.123 | -0.104 |

The smoother nanopore-array response is most directly explained by ensemble averaging. In a single pore, nucleation, precipitate growth, surface-charge changes, blockage, and reopening occur within one confined pathway, so a small local difference can shift the entire device between blocked and open states. In the 3 × 3 array, the measured current sums nine local responses, allowing conductive pathways to remain while other pores precipitate or reopen. For independent pores with comparable means and variances, the relative fluctuation of the summed current scales as $N^{-1/2}$ and would be approximately threefold lower for N = 9. This scaling follows from the independent-pore assumption; related ensemble behavior has been discussed in multipore memristive systems[38]. It is an idealized benchmark because individual pore currents and replicate fluctuation statistics were not measured here. Neighboring pores may also interact through overlapping electric-field and access-resistance regions, which can make array conductance subadditive. Current-dependent voltage loss in a shared access region could moderate the effective pore voltage, but local voltages and pore-resolved currents were not measured. This mechanism therefore remains speculative[20-23].

Overlapping concentration-polarization zones could likewise couple the local $Ca^{2+}$ and phosphate activities. Given the relatively high phosphate concentrations and the absence of pore-resolved concentration profiles, this is unlikely to be the primary explanation here. The data most directly support ensemble averaging; access resistance and concentration coupling remain secondary possibilities that require dedicated measurements[39]. Scan-rate dependence links the I–V hysteresis to finite ion-transport and precipitation times. At both pH 8.5 and 100 mM $Na_2HPO_4$, nanopores array $A_{norm}$ increased over 0.04–0.12 V $s^{-1}$, whereas the single pore peaked at 0.06 V $s^{-1}$. We also evaluated the capacitive contribution from the branch-current separation, using $C_{app} = |I_{increasing} - I_{decreasing}|/(2\nu)$ at ±250 mV (Supplementary Fig. S9). For the nanopores array positive-bias closed state, $C_{app}$ remained within 16.6–18.0 nF at pH 8.5 and 14.6–19.7 nF at 100 mM $Na_2HPO_4$, consistent with a reproducible capacitive component. In contrast, negative-bias $C_{app}$ varied from 132.4 to 913.0 nF at pH 8.5 and from 12.7 to 345.2 nF at 100 mM. The single pore also showed strongly scan-rate-dependent apparent values. The negative-bias open-state current therefore cannot be described as purely capacitive; it contains substantial reaction- and state-dependent contributions. $A_{norm}$ should accordingly be interpreted as a composite measure of capacitive and reaction-mediated hysteresis. With only three scan rates, these data do not support a quantitative kinetic model.[6, 7, 18, 20, 37]. pH and phosphate concentration alter speciation, supersaturation, nucleation probability, and the persistence of residual solid[28-30, 33, 34]. Increasing reactant supply can widen the response window, block a single pore too strongly, or leave nuclei that accelerate the next event. The single pore makes these local transitions visible; the nanopores array converts them into a device-level signal only when enough parallel pathways remain conductive and recoverable. The term stability must therefore be used narrowly. The present data support smoother I–V curves and more consistent pulse-endpoint polarity for the array at 100 mM, not long-term endurance or device-to-device reproducibility. Likewise, positive and negative ΔI values denote endpoint current shifts; they should not be renamed synaptic potentiation and depression without a reproducible polarity protocol, retention data, and cycle statistics. Prior precipitation-gated and geometry-programmed devices supported stability claims with reversible cycling, independent pulse cycles, retention measurements, or reset tests[30, 32, 37]. Those validation layers are not present in the current dataset. Our working mechanism is voltage-controlled precipitation gating. Under the polarity convention used in this study, negative bias corresponds to the open state and positive bias to the closed state. Positive bias is therefore associated with formation or persistence of a calcium-phosphate blockage that narrows the conducting cross-section, whereas reversal to negative bias promotes reopening through redistribution of reactants and precipitate dissolution. The exact phase and the relative roles of dissolution and phosphate speciation remain unresolved. In the array, these events occur in several parallel pores and are summed electrically. This model explains the observed polarity, geometry, and scan-rate dependence, but remains an inference from current rather than direct phase-resolved imaging[32,40]. The ϕ = 150 nm pore provides an auxiliary geometry control (Supplementary Figs. S1 and S2). The primary geometry comparison therefore remains limited to the ϕ = 250 nm single pore and 3 × 3 nanopores array, which share the same nominal individual pore size.

# Conclusion

We built an asymmetric $CaCl_2$/phosphate electrolyte system across $SiN_x$ solid-state nanopores. The experiments reveal the coupled roles of pH, phosphate supply, and pore geometry in precipitation-gated memory. The normalized hysteresis area of the pore array decreased at high pH. The ϕ = 250 nm single pore showed stronger local and nonmonotonic behavior. Thus, array averaging and confined single-pore reactions produced distinct responses. The concentration series also showed a response increase above about 10 mM in the ϕ = 250 nm pore. Within the tested conditions, sufficient reactant supply was one requirement for precipitation-gated switching. Pulse sampling showed that voltage-induced conductance states remained readable at a fixed sampling voltage. Electrolyte conditions, pulse amplitude, and pore geometry jointly controlled $\Delta I$. At 100 mM phosphate, the nanopore array produced a more consistent positive endpoint response across the tested programming windows. Excess concentration or strong pulses may cause overdeposition, local blockage, and incomplete recovery. These effects can weaken or reverse the readout. The Supporting Information reports the auxiliary ϕ = 150 nm control separately from the primary device comparison. The results support a phosphate precipitation-gating mechanism. Voltage concentrates $Ca^{2+}$ and phosphate ions inside the nanopore and drives precipitation and redissolution. These processes alter the conductive cross-section and interfacial charge, which creates readable conductance memory. This mechanism guides concentration-window selection, pore-size design, and array construction for chemically tunable nanofluidic memristors. Future in situ measurements and cycling tests should separate reversible precipitation, irreversible blockage, and surface-state drift[31].

## Materials and Methods

### Nanopore fabrication

We fabricated $SiN_x$ membrane chips with standard microelectromechanical systems (MEMS) processes. We used a commercial 500 μm silicon substrate coated on both sides with 100 nm LPCVD $SiN_x$. Ultraviolet photolithography with S1813 defined square $SiN_x$ windows on one side. The pattern period was 5 mm, and each window was about 850 μm wide. We removed the exposed $SiN_x$ by $CHF_3/O_2$ reactive ion etching (RIE). We then cleaned the wafer with acetone and oxygen plasma. Next, we immersed the wafer in 32% KOH at 90 °C for several hours. This step removed silicon beneath each window and formed a free-standing $SiN_x$ membrane. Finally, we milled nanopores with a diameter of about 250 nm with an FEI 650 focused ion beam (FIB). We used an 18 pA beam current and a 30 kV acceleration voltage.

## Ionic current measurements

We performed electrical measurements in a custom microfluidic cell fitted with Ag/AgCl electrodes. An Elements srl electrical reader recorded the ionic current. The cis Ag/AgCl electrode was held at 0 V, and the command voltage was applied to the trans electrode; voltage is therefore defined as $V = V_{trans} - V_{cis}$. Under this convention, negative bias is the open state and positive bias is the closed state. We ran at least three voltage sweeps for each condition. We used the arithmetic mean current to evaluate reaction-mediated ion transport inside the pore. We used all salts without further purification. We purchased $CaCl_2$ (>95.0% purity) and $Na_2HPO_4$ (>99.0% purity) from Sigma-Aldrich.

We used one cleaning protocol to prevent cross-contamination and unintended precipitation during solution changes. First, we removed the old solution from the microfluidic cell with a pipette. We then flushed the inlet and outlet channels at least five times with Milli-Q water. Finally, we added the new salt solution. For pH experiments, we adjusted phosphate-buffered saline (PBS) to pH 5.5, 6.5, 7.5, 8.5, or 9.5 with hydrochloric acid or sodium hydroxide. For concentration experiments, we used 1, 5, 10, 50, or 100 mM $Na_2HPO_4$, adjust the pH with dilute hydrochloric acid. We plasma-cleaned each chip for 300 s before testing. We added $CaCl_2$ (2 M) to the upper cis chamber and PBS or $Na_2HPO_4$ to the lower trans chamber.

## Hysteresis-loop measurements

We used I–V curves to measure history dependence during voltage sweeps. Each curve was split into forward and reverse branches. The hysteresis area, $A_{hys}$, was calculated between the branches over their common voltage interval. For cross-condition comparison, we defined $A_{norm} = A_{hys}/[(V_{max} - V_{min})(I_{max} - I_{min})]$. This dimensionless metric scales the loop area by the measured voltage and current spans, whereas $A_{hys}$ retains the absolute response magnitude. For each bias regime, $G_{d,max}^{\pm}$ is the maximum local differential conductance, *dI/dV*. $G_{fit}^{\pm}$ is the slope of a linear fit to the high-bias segment of the branch-averaged I–V curve. We estimated the apparent capacitance at ±250 mV as $C_{app} = |I_{increasing} - I_{decreasing}|/(2\nu)$, where ν is the scan rate. The curves presented in the main text for the single pore system have been corrected by the 0 V capacitance taking into consideration the measurement from a single $SiN_x$ membrane without a pore a substracting the current level.

### Pulse-sampling measurements

We used pulse tests to examine temporal accumulation of the conductance state. The nanopore array was programmed at ±40–80 mV and read at +20 mV; larger voltages drove the summed array current beyond the reader's 200 nA range. The 250 nm single pore was programmed at ±300–500 mV and read at +200 mV to provide a clear dynamic range. The auxiliary 150 nm single pore followed the lower-voltage array protocol and was read at +20 mV. The pulse duration was 100 ms for the nanopore array and the 150 nm control, and the baseline duration was 500 ms.

### Acknowledgements

We are grateful for continued financial support from National Key Research and Development Project of China (No. 2023YFF0613603), National Natural Science Foundation of China (No. 22202167), Provincial Science and Technology Plan Project: Micro and Nano Preparation and Photoelectronic Detection (No. 03014/226063). D. G. thank the HORIZON MSCA DN-2022: DYNAMO, grant Agreement 101072818. A part of this work was supported by the Japan Society for the Promotion of Science (JSPS) KAKENHI Grant Numbers 22H01926, 22H01410, and 24K21715. M. Tsutsui acknowledges support from Kansai Research Foundation for Technology Promotion.